\documentclass[12pt]{article}
\usepackage{times}
\usepackage{graphics}
\usepackage{graphicx}
\usepackage[active]{srcltx}
\usepackage{cite}

\newcommand{\n}{\mbox{\boldmath $n$}}
\newcommand{\nq}{\mbox{\boldmath $n_{q}$}}
\newcommand{\q}{\mbox{\boldmath $q$}}
\newcommand{\kk}{\mbox{\boldmath $k$}}
\newcommand{\ko}{\mbox{\boldmath $k_{0}$}}
\newcommand{\si}{\mbox{\boldmath $\sigma$}}
\newcommand{\rr}{\mbox{\boldmath $r$}}

\begin{document}
\large
\date{ }

\begin{center}
{\Large Possible polarized neutron-nucleus scattering search for a new spin-dependent nucleon-nucleon coupling in a Fm range}

\vskip 0.4cm

Yu. N. Pokotilovski \footnote{e-mail: pokot@nf.jinr.ru}

\vskip 0.4cm
            Joint Institute for Nuclear Research\\
              141980 Dubna, Moscow region, Russia\\
\vskip 0.4cm

{\bf Abstract\\}

\begin{minipage}{130mm}

\vskip 0.4cm
 An experimental opportunity is presented for the future to measure possible
P- and T-non-invariant axion-like interaction between nucleons in a Fm range.
 This interaction may be searched for in the measurement spin-dependent 
asymmetry of scattering of polarized neutrons in a keV-MeV energy range by 
heavy nuclei.

\end{minipage}
\end{center}

\vskip 0.3cm


\vskip 0.2cm


\vskip 0.6cm

\section{Introduction}
 Possible new hypothetical interactions coupling mass to particle spin were
discussed in a number of papers \cite {Lei,Moh,Hill}.
 At the same time, there are serious theoretical arguments that there may exist
scalar or pseudoscalar, vector or pseudovector, weakly interacting bosons.
 Generally the proposed models can not predict the masses of these particles
and their coupling to nucleons, leptons, and photons.

 The most popular and actively discussed possibility for the existence of new
light particles is a pseudoscalar boson - axion \cite{ax}.
 Its existence provides attractive solution of the strong CP problem \cite{PQ}.

 The original "standard" axion model \cite{ax} gave strict predictions for the
mass of axion and coupling constants between axions and other particles but was
disproved by experiments.

 New theoretical models of "invisible" axions: "hadronic" or KSVZ \cite{KSVZ}
and DFSZ \cite{DFSZ} protected axion from existing experimental constraints
suppressing their interaction with matter, but retaining at the same time
possibility of solution of the strong CP problem.

 In all these models of pseudo-Goldstone boson, the masses and coupling constants
are determined by one constant $f_{a}$ -- the scale of the global $U(1)_{P-Q}$
-- Peccei-Quinn symmetry breaking, and axion may have mass in a very large
range:  $(10^{-12}<m_{a}<10^{6})$\,eV.
 Current algebra relates the masses of the axion and neutral pion:
$m_{a}f_{a}=m_{\pi}f_{\pi}\sqrt{z}/(z+1)$, where $z=m_{u}/m_{d}=0.56$,
$f_{\pi}\approx 93$\,MeV, $m_{\pi}=135$\,MeV, so that
$m_{a}$\,(eV)$\approx 6\times 10^{6}/f_{a}$\,(GeV).
 The coupling of the axion with a fermion with mass $m_{f}$ can be in general
expressed as $g_{aff}=C_{f}m_{f}/f_{a}$, where $C_{f}$ is the model dependent
factor \cite{Kim,PDG}.

 Numerous nuclear transition, weak decays, reactor, and beam-dump experiments
have placed limits on the axion mass and coupling constants.
 The most stringent limits, especially from the lower side of the axion mass
range, have been set on the existence of axion using astrophysical and
cosmological arguments \cite{Kim,PDG,win}.
 Most recent constraints limit the axion mass to rather low values:
$(10^{-5}<m_{a}<10^{-3})$\,eV with very small coupling constants to quarks and
photon.

 Axion is still one of the candidates for the cold dark matter of
the Universe \cite{dark}.

 These limits are more strong than are reached in laboratory experiments.
 Nevertheless it is of interest to try to constrain the axion using laboratory
means because interpretation of laboratory experiments depend on less number of
assumptions than the constraints inferred from astrophysical and cosmological
observations and calculations.

 On the other hand the direct experimental constraints on the axion-like
interaction may be useful for limiting more general class of bosons and
spin-dependent interactions irrespective to any particular theoretical model.

 Other models proposed to solve the strong CP problem come from the idea of
mirror world \cite{Berr} and super-symmetry \cite{SUSY}.
 These models allow for existence of pseudoscalars with much larger masses.

 Various unifications of Standard Model predict new vector and pseudovector
weakly interacting bosons \cite{sp1,Hol}.

 New possible experimental schemes for the search for these particles with
masses from a few Mev are discussed \cite{Ess,Bjor} and old experimental data
are reprocessed \cite{Blu}.

 On the other hand some experimental evidence of existence of scalar or
pseudoscalar particles in the mass range from one to dozens MeV appeared
last years in a number of publications.

 Measurements of electron-positron pair production by high energy particles in
emulsions yielded significant excess of events with angles between electron and
positron tracks larger than predicted from QED calculations of external pair
production by photons (see recent reanalysis and review article \cite{deB} and
references therein).
 The deviation was interpreted as decays of neutral bosons with lifetimes
between $10^{-16}$ and $10^{-14}$ s.
 Probable mass of these particles was found to be between 1.5 and 30 MeV.
 Additional evidence in favor of existence of neutral bosons in this mass
range comes from the investigation of internal pair conversion in decays of
excited states of light nuclei (references can be found in \cite{deB}).

 Particles with mass in a MeV range are considered \cite{Cos1,Cos2} as possible
dark matter candidates in connection with the observed 511 keV gamma-ray line
from the galactic center \cite{511}.

 Berezhiani and Drago \cite{Ber} in attempt to resolve puzzles of giant gamma-
ray bursts assumed the existence of pseudo-Goldstone bosons heavier than
already constrained light axion.
 They found that pseudoscalars in a mass range 0.3 MeV$<m_{a}<$\,few MeV
can provide mechanism for the cosmic gamma-ray bursts.
  Their approach was purely phenomenological, considering masses $m_{a}$ and
couplings $g_{a\gamma}$, $g_{aN}$, $g_{ae}$ as independent.
 These "large" masses are not excluded by laboratory experiments or
astrophysical calculations based upon stellar evolution models.

 Laboratory search for weakly interacting neutral pseudoscalar particles in
beam-dump  experiments (for review see \cite{Kim}) is based on production and
subsequent decay of these particles:
$\pi^{+}\rightarrow e^{+}+\nu_{e}+a \, (a\rightarrow e^{+}e^{-})$,
$K^{+}\rightarrow\pi^{+}+a$\,(invisible),
$B^{+}\rightarrow K^{+}+a$\,(invisible),
$J/\psi\rightarrow \gamma+a$\,(invisible),
$\Upsilon\rightarrow \gamma+a$\,(invisible)
and includes assumptions on decay properties of these particles.

 Very recently Gninenko \cite{Gni} considered the experimental data on the
radiative decays of neutral $\pi$-mesons in a hypothetical mode
$\pi^{0}\rightarrow\gamma X$, where $X$ is a new boson in a MeV mass range to
constrain the coupling of $X$ with ordinary matter.

 Decays $\phi\rightarrow\eta U$ with the following decay
$U\rightarrow e^{+}e^{-}$, where $U$ is pseudovector boson with mass
$m_{U}>5$ MeV were used in \cite{phi} to limit interaction strength of these
particles.

 Searches of axions with an energy of 5.5 MeV produced in the $p(d,^{3}He)a$
reaction in the Sun were performed in \cite{pd,pd1}.
 The range of axion mass has been extended in these works to 5 MeV.
 The axion detection scheme used in these studies includes
Compton$\rightarrow$axion conversion $a+e\rightarrow e+\gamma$, the
axio-electric effect $a+e+Z\rightarrow e+Z$ and axion$\rightarrow$photon
conversion $a+Z\rightarrow \gamma+Z$ (Primakoff process).

 We consider here possible low-energy (below 1 MeV) polarized neutron-nucleus
scattering experiment sensitive to the axion-like interaction in the range
$\lambda=\hbar/(m_{a}c)\div(10^{-13}-10^{-9})$\,cm.

 In the nucleon-nucleus scattering approach no assumption is made
on decay properties of particles or their interaction with leptons and
photons.

\section{Monopole-dipole interaction}
 A P- and T-odd  monopole-dipole interaction potential between spin and matter
(polarized and unpolarized nucleons) has a form \cite{Mood}:
\begin{equation}\label{1}
U({\bf r})=
(\si\cdot\n)\frac{g_{s}g_{p}\hbar^2}{8\pi m_{n}} \Biggl(\frac{1}{\lambda r}+
\frac{1}{r^{2}}\Biggr)e^{-r/\lambda},
\end{equation}
where $g_{s}$ and $g_{p}$ are dimensionless coupling constants of the scalar
and pseudoscalar vertices (unpolarized and polarized particles), $m_{n}$ is the
nucleon mass at the polarized vertex, $\si$ is vector of the Pauli matrices
related to the nucleon spin, $r$ is the distance between the nucleons,
$\lambda=\hbar/m_{a}c$ is the range of the force, $m_{a}$ is the axion mass,
and $\n=\rr/r$ is the unit vector.

 In the notations of the general classification of all possible interactions
between two nonrelativistic spin 1/2 particles \cite{Dob} this potential is of
the type $V_{9,10}$.

 The neutron-nucleus scattering amplitude following from the interaction of 
Eq. (1) is
\begin{equation}\label{2}
f_{a}(\q)=-ig_{s}g_{p}N\frac{\si\cdot\nq}{4\pi}
\frac{q\lambda^{2}}{1+(q\lambda)^{2}},
\end{equation}
where $\q$ is the neutron scattering vector, $\nq=\q/q$ is the unit vector,
$q=2\,k\,sin(\theta/2)$, the neutron wave vector in the center of mass system
$k (F^{-1})=2.1968\times 10^{-4}\frac{A}{A+1}E_{n}^{1/2}$, the neutron energy
$E_{n}$ in the lab. system in eV, and $N$ is the nucleon number in the nucleus.

 This amplitude is maximal at $q=1/\lambda$ and is
\begin{equation}\label{3}
f_{a}(\lambda)=\frac{\pm ig_{s}g_{p}N\lambda}{8\pi},
\end{equation}
the sign depending on the sign of the scalar product $\si\cdot\nq$.

 The observable in a polarized neutron-nucleus scattering test of monopole-
dipole interaction is the left-right or forward-backward spin-dependent
asymmetry of scattering of transversely or, respectively, longitudinally
polarized neutrons.

 Neutrons with energies in the range ($10<E_{n}<10^{6}$) eV correspond to the 
$\lambda$ range in Eqs. (2) and (3): ($3<\lambda<10^{3}$) F, and $m_{a}$ range 
($0.2<m_{a}<70$) MeV.

 The low-energy neutron-nucleus scattering amplitude is
\begin{equation}
f(q)=f_{0}(q)+f_{1}(q)+f_{\tiny{\mbox{Sch}}}(q)+f_{n-e}(q)+f_{p}(q)+f_{a}(q),
\end{equation}
where $f_{0}(q)$ is the amplitude of $s$-scattering, $f_{1}(q)$ is the
amplitude of $p$-scattering, $f_{\tiny{\mbox{Sch}}}$ is the amplitude
of electromagnetic (Schwinger) scattering of the neutron magnetic moment
in the Coulomb field of the nucleus, $f_{n-e}$ is the amplitude of the neutron
scattering by atomic electrons, $f_{p}$ is the amplitude of neutron-nucleus
scattering due to the neutron electric polarizability, $f_{a}$ is the amplitude of a
hypothetical long-range spin-dependent neutron-nucleus interaction.
 Higher neutron angular momenta were neglected in this expression.

 As is known small scattering amplitudes $f_{n-e}$ and $f_{p}$ are determined
experimentally through their interference with the real part of the
neutron-nucleus scattering amplitude \cite{Alex}.

 Our approach to search for the contribution of axion-like interaction to the
polarized neutron-nucleus scattering consists in the interference of imaginary
$f_{a}$ amplitude of Eqs. (2) and (3) with imaginary part of the
neutron-nucleus scattering amplitude.

 In what follows we omit spin-independent small terms $f_{n-e}$ and $f_{p}$.
 The Schwinger amplitude is imaginary and does not depend on energy \cite{Schw}:
\begin{equation}
f_{\tiny{\mbox{Sch}}}=i\frac{\mu_{n}Ze^{2}}{2m_{n}c^{2}}\frac{\Bigl(\si[\kk\ko]\Bigr)}{kk_{0}}ctg(\frac{\phi}{2}).
\end{equation}
 Here Z is the nuclear charge, $\ko$ and $\kk$ are the incident and final
neutron wave vectors,  $\mu_{0}$ is the neutron magnetic moment in nuclear
magnetons $e\hbar/(2mc)$, and $\phi$ is the neutron scattering angle.

 Interference of the Schwinger amplitude with imaginary part of the
neutron-nucleus scattering amplitude can imitate effect of the axion-like interaction in a measurement of
left-right spin-dependent scattering asymmetry of transversely polarized neutrons at non-perfect symmetry of the
neutron detectors in respect to the plane determined by the vectors $\ko$ and $\si$.
 In the measurement of spin-dependence of forward or backward neutron-nucleus
scattering of strongly longitudinally polarized neutrons the contribution of
the Schwinger scattering to spin-dependent effect is zero.

 The scattering cross-section has the form
\begin{eqnarray}
\frac{d\sigma^{\pm}}{d\Omega}\approx
\Bigl|Re f_{0}+i\cdot Im f_{0}+Re f_{1}+i\cdot Im f_{1}
\pm i\cdot|f_{a}|\Bigr|^{2}\approx \nonumber\\
\approx (Re f_{0}+Re f_{1})^{2}+(Im f_{0}+Im f_{1})^{2}
\pm 2\,(Imf_{0}+Im f_{1})\,|f_{a}| \nonumber\\
\approx \frac{\sigma_{tot}}{4\pi}\pm 2\,(Im f_{0}+Im f_{1})\,|f_{a}|
=\frac{\sigma_{tot}}{4\pi}\pm 2\,Im f\,|f_{a}|.
\end{eqnarray}
 This expression is approximate because exact geometry of the measurement is not specified here.
 Scattering-angle-dependent contribution of p-scattering amplitude $f_{1}$ is a function of the detectors
position: whether transverse or longitudinal spin-dependent scattering asymmetry is measured of,
respectively, transversely or longitudinally polarized neutrons and the solid angle of detectors.

\section{Spin-dependent scattering asymmetry}
 From the spin-dependent scattering asymmetry determined as
\begin{displaymath}
A=\frac{d\sigma^{+}/d\Omega-d\sigma^{-}/d\Omega}{d\sigma^{+}/d\Omega+d\sigma^{-}/d\Omega},
\end{displaymath}
we have
\begin{equation}
A=\frac{d\sigma^{+}/d\Omega-d\sigma^{-}/d\Omega} {d\sigma^{+}/d\Omega+d\sigma^{-}/d\Omega}= \frac{8\pi Im
f\cdot|f_{a}|}{\sigma_{tot}}=g_{s}g_{p}\frac{Im f\cdot  N\lambda}{\sigma_{tot}}=g_{s}g_{p}B\lambda,
\end{equation}
where $B=Im f\cdot N/\sigma_{tot}$ is the sensitivity of the measurement, one can find the product of the
coupling constants:
\begin{equation}
g_{s}g_{p}=A\frac{\sigma_{tot}}{Im f\cdot N\lambda}=\frac{A}{B\lambda}.
\end{equation}
 The order of magnitude estimate gives
$g_{s}g_{p}=A/\lambda\,(Fm)$, if $\sigma_{tot}$=10 b, $Im f$=5 Fm, N=200.

 To calculate the real and imaginary parts of the neutron-nucleus scattering
amplitudes we use the following formalism:
\begin{equation}
f_{0}=\frac{i}{2k}(1-S_{0}),
\end{equation}
where S-matrix
\begin{equation}
S_{0}=e^{2i\delta_{0}}\Biggl(1-\sum_{j}\frac{ig_{j}\Gamma_{nj}} {(E-E_{j})+i\Gamma_{j}/2}\Biggr).
\end{equation}
 $\Gamma_{nj}=\Gamma_{nj}(E_{j})k/k_{j}$,
$\Gamma_{j}=\Gamma_{nj}+\Gamma_{\gamma j}$, and $g_{j}$ are the
neutron widths, the total widths, and the statistical weights of $j$-th $s$-resonance.

 Upon introducing
\begin{equation}
\sum\nolimits_{(1)}=\sum_{j}\frac{2}{k_{j}}\frac{g_{j}\Gamma_{nj}
(E-E_{j})}{4(E-E_{j})^{2}+\Gamma_{j}^{2}},
\end{equation}
and
\begin{equation}
\sum\nolimits_{(2)}=\sum_{j}\frac{1}{k_{j}}\frac{g_{j}
\Gamma_{nj}\Gamma_{j}}{4(E-E_{j})^{2}+\Gamma_{j}^{2}},
\end{equation}
the real and the imaginary parts of the nuclear scattering amplitude are
\begin{equation}
{\mbox{Re}}f_{0}=\frac{{\mbox{sin}}(2\delta_{0})}{2k}-
{\mbox{cos}}(2\delta_{0})\sum\nolimits_{(1)}-
{\mbox{sin}}(2\delta_{0})\sum\nolimits_{(2)}
\end{equation} and
\begin{equation}
{\mbox{Im}}f_{0}=\frac{{\mbox{sin}}^{2}\delta_{0}}{k}+
{\mbox{cos}}(2\delta_{0})\sum\nolimits_{(2)}-{\mbox{sin}}(2\delta_{0})
\sum\nolimits_{(1)}.
\end{equation}

The $p$-scattering amplitude is
\begin{equation}
f_{1}=\frac{i}{2k}(1-S_{1}),
\end{equation}
\begin{equation}
S_{1}=e^{2i\delta_{1}}\Biggl(1-\sum_{j}\frac{ig_{j}\Gamma_{nj}}
{(E-E_{j})+i\Gamma_{j}/2}\Biggr).
\end{equation}
where $\Gamma_{j}=\Gamma_{nj}+\Gamma_{\gamma j}$, and $\Gamma_{j},\,
\Gamma_{nj},\, \Gamma_{\gamma j}$, and $g_{j}$ are the total
width, the neutron width, the gamma-width and the statistical
weight of $j$-th $p$-resonance.
\begin{equation}
\Gamma_{nj}=\frac{k}{k_{j}}\Gamma_{nj}(E_{j})\frac{v_{1j}}{v_{0j}},
\end{equation}
where
\begin{equation}
v_{0j}=\frac{(k_{j}R)^{2}}{1+(k_{j}R)^{2}},\qquad
v_{1j}=\frac{(kR)^{2}}{1+(kR)^{2}}.
\end{equation}
The phase of $p$-scattering
\begin{equation}
\delta_{1}=-kR+{\mbox{arctg}}(kR)+
{\mbox{arcsin}}\Biggl[\frac{k}{3}\cdot\frac{(kR)^{2}}{1+(kR)^{2}}
\bigl(R-R^{'}_{1}\bigr)\Biggr],
\end{equation}
where $R$ is the channel radius, $R^{'}_{1}$ is the $p$-wave scattering
radius.
The summation in Eqs. (10-16) is performed over all known $s$- and 
$p$-resonances \cite{Mug}.

 Due to the presence of numerous resonances in the neutron-nucleus scattering and interference of potential and
resonance amplitudes the sensitivity $B(E_{n})$ fluctuates reaching significant values in narrow energy intervals,
where the total neutron cross section is small.
 Figs. 1 and 2 show the calculated s- and p-scattering amplitudes for $^{207}$Pb,
and Fig. 3 illustrates the sensitivity factor $B(E_{n})$ for this nucleus as a function
of energy.

 Strong fluctuation of the sensitivity in accordance with fluctuations of imaginary part of
the nuutron-nucleus scattering amplitude gives opportunity to search for the
spin-dependent scattering asymmetry in time-of-flight mode without special measurement of the background effect.
 With B=10 Fm$^{-1}$ and measured asymmetry $A=10^{-4}$ the limit on $g_{s}g_{p}$ may be obtained
at the level $10^{-6}$ at $\lambda=10$\, Fm.

 Weak interactions violating P and T symmetries may imitate discussed effects
of spin-asymmetries in polarized neutron - unpolarized nucleus scattering.
 Estimates of these effects are contained in \cite{Bar,Lyu} (see also the
recent book of Barabanov \cite{Barb}).
 It was shown that even in p-resonances at low energies, where weak interaction
effects are largest due to mixture of p- and s- resonances in weak interactions,
the value of asymmetries is small: $\sim 10^{-8}-10^{-7}$.
 Far from p-resonances the effect should be orders of magnitude lower.

 The neutron crystal-diffraction method was proposed recently in \cite{Vor}
to search for the short-range spin-dependent axion-like nucleon-nucleon
interactions.
 It is based on the measurement of spin rotation of thermal neutrons passing
through a non-centrosymmetric crystal.
 Their constraints on axion-like interactions followed from the results of 
their test search for the neutron EDM performed with the same method.
 They obtained the most stringent current limits: the product of the scalar and pseudoscalar
constants of the axion-nucleon interactions $g_{s}g_{p}\le 10^{-12}$ for
the interaction range $\lambda\ge 10^{-8}$\,cm and
$g_{s}g_{p}\le 3\times 10^{-29}/\lambda^{2}$ for $\lambda\le 10^{-9}$\,cm
with expected significant progress in future measurements.

\newpage
\newpage
\begin{figure}
\begin{center}
\resizebox{18cm}{12cm}{\includegraphics[width=\columnwidth]{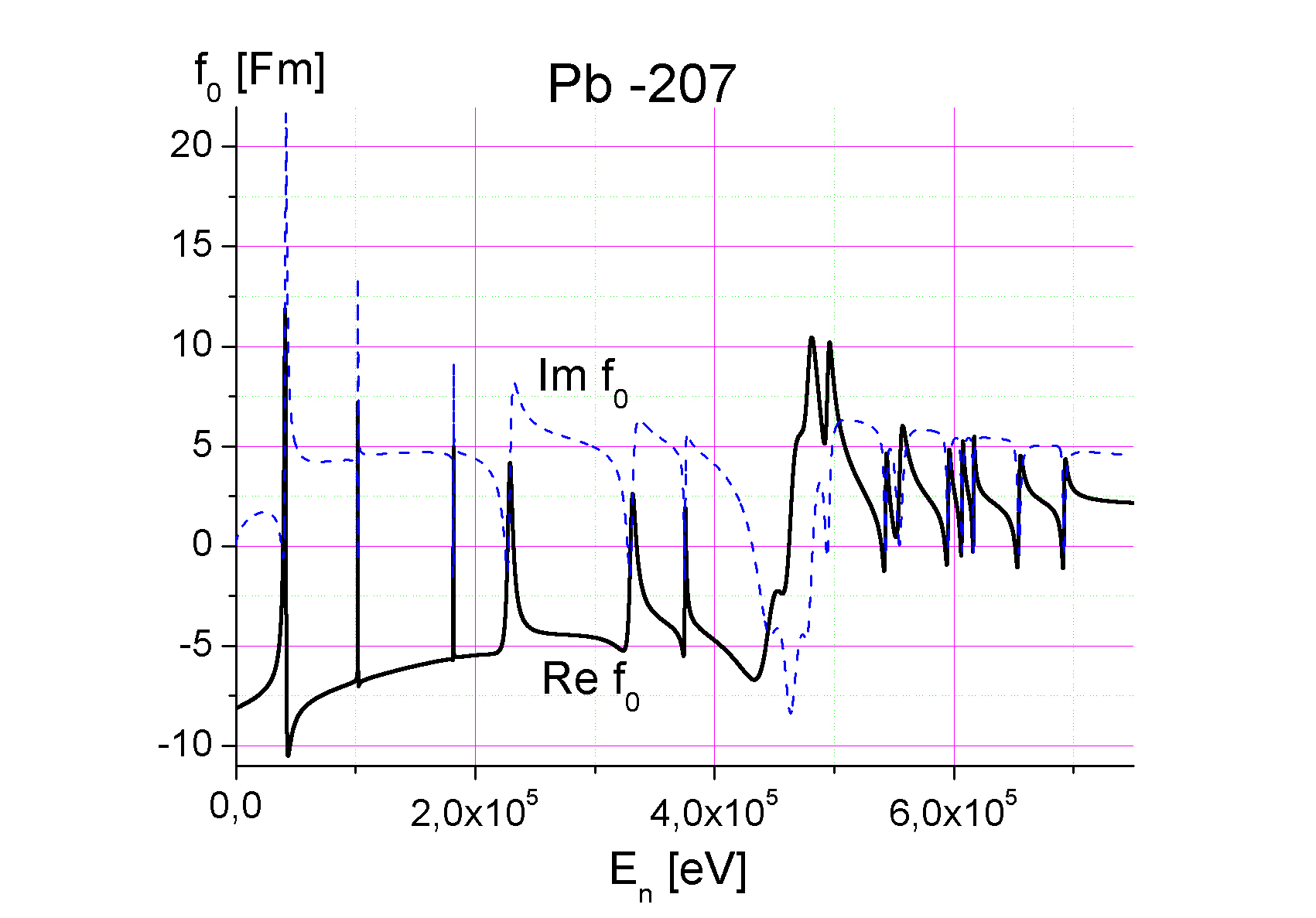}}
\end{center}
\caption{Calculated neutron s-scattering amplitude for $^{207}$Pb.}
\end{figure}

\newpage
\begin{figure}
\begin{center}
\resizebox{18cm}{12cm}{\includegraphics[width=\columnwidth]{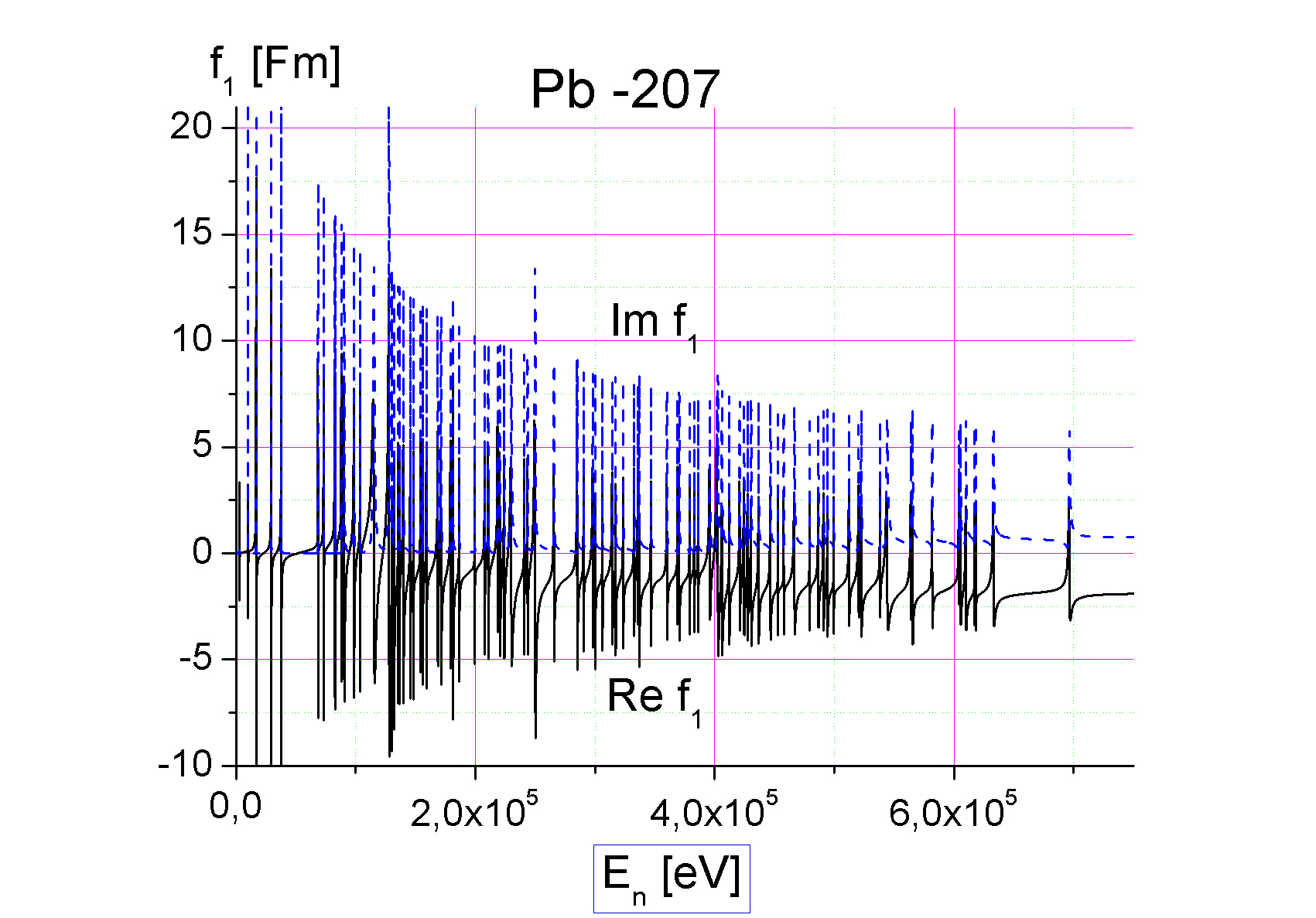}}
\end{center}
\caption{Calculated neutron p-scattering amplitude for $^{207}$Pb.}
\end{figure}

\newpage
\begin{figure}
\begin{center}
\resizebox{18cm}{12cm}{\includegraphics[width=\columnwidth]{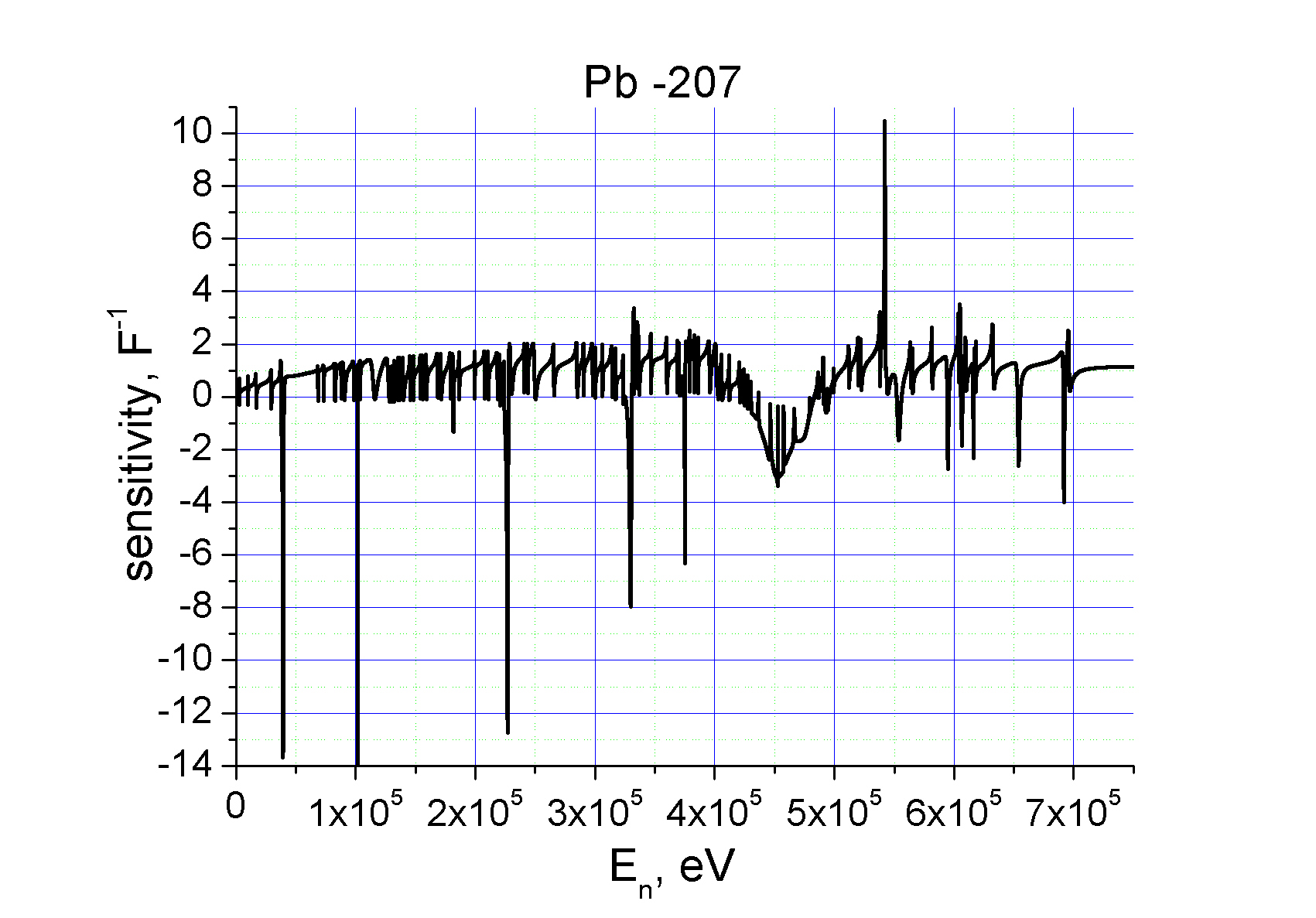}}
\end{center}
\caption{Calculated sensitivity $B(E_{n})$ for $^{207}$Pb.}
\end{figure}


\begin{thebibliography}{92}
\bibitem{Lei}
J. Leitner and S. Okubo, Phys. Rev. {\bf 136} (1964) B1542.

\bibitem{Moh}
D. Chang, R.N. Mohapatra, and S. Nussinov, Phys. Rev. Lett. {\bf 55} (1985)
2835.

\bibitem{Hill}
C. T. Hill and G. G. Ross, Nucl. Phys. {\bf B 311} (1988) 253.

\bibitem{ax}
S. Weinberg, Phys. Rev. Lett. {\bf 40} (1978) 223; F. Wilczek, Phys. Rev. Lett. {\bf 40} (1978) 279.

\bibitem{PQ}
R.D. Peccei and H.R. Quinn, Phys. Rev. Lett. {\bf 38} (1977) 1440;
Phys. Rev. {\bf D11} (1977) 1791.

\bibitem{KSVZ}
J.E. Kim, Phys. Rev. Lett., {\bf 43} (1979) 103;

M.A. Shifman, A.I. Vainstein, V.I. Zakharov, Nucl. Phys., {\bf B166} (1980)
493.

\bibitem{DFSZ}
A. Zhitnitsky, Yad. Phys., {\bf  31} (1980) 497; [Sov. Journ. Nucl. Phs. {\bf 31} (1980) 260].

M. Dine, F. Fischler, and M. Srednicki, Phys. Lett., {\bf B104} (1981) 199.

\bibitem{Kim}
J.E. Kim, Phys. Rep. {\bf 150} (1987) 1.

\bibitem{PDG}
C. Hangmann, H. Murayama, G.G. Raffelt, L.J. Rosenberg, and K. van Bibber
(Particle Data Group), Journ. Phys. {\bf G37} (2010) 496.

\bibitem{win}
M. S. Turner, Phys. Rep. {\bf 197} (1990) 67.

\bibitem{dark}
P. Sikivie, Phys. Rev. Lett. {\bf 51} (1983) 1415; R. Bradley,
J. Clarke, D. Kinion et al. Rev. Mod. Phys. {\bf 75} (2003) 777.

\bibitem{Berr}
Z. Berezhiani, L. Gianfagna, M. Gianotti,  Phys. Lett., {\bf B500} (2001) 286.

\bibitem{SUSY}
L.J. Hall and T. Watari, Phys. Rev. {\bf D70} (2004) 115001.

\bibitem{sp1}
P. Fayet, Phys. Lett. {\bf 84B} (1979) 416; {\bf 95B} (1980) 285;
{\bf 96B} (1980) 83.

P. Fayet and M. Mezard, Phys. Lett. {\bf B104} (1981) 226.

P. Fayet, Nucl. Phys. {\bf B187} (1981) 184; {\bf B237} (1984) 367;
Phys. Lett. {\bf B171} (1986) 261; {\bf B172} (1986) 363.

P. Fayet, Nucl. Phys. {\bf B347} (1990) 743;
Class. Quant. Grav. {\bf 13} (1996) A19;
Phys. Rev. {\bf D75} (2007) 115017; Phys. Lett. {\bf B675} (2009) 267.

\bibitem{Hol}
B. Holdom, Phys. Lett. {\bf B166} (1986) 196.

\bibitem{Ess}
R. Essig, R. Harhik, J. Kaplan, and N. Toro, arXive: 1008.0636 [hep-ph].

\bibitem{Bjor}
J.D. Bjorken, R. Essig, P. Schuster, and N. Toro, Phys. Rev. {\bf D80} (2009)
075018.

\bibitem{Blu}
J. Bl\"umlein, J. Brunner, Phys. Lett., {\bf B701} (2011) 155.

\bibitem{deB}
F.W.N. de Boer, C.A. Fields, Int. Journ. Mod. Phys. {\bf B20} (2011) 1787.

\bibitem{Cos1}
C. Boehm, P. Fayet, Phys. Lett. {\bf B683} (2004) 219.

\bibitem{Cos2}
C. Boehm, D. Hooper, J. Silk et al.,   Phys. Rev. Lett. {\bf 92} (2004) 101301.

\bibitem{511}
P. Jean et al., Astron. Astrophys., {\bf 407} (2003) L55.

\bibitem{Ber}
Z. Berezhiani, A.Drago, Phys. Lett. {\bf B573} (2000) 281.

\bibitem{Gni}
S.N. Gninenko, arXiv:1112.5438 [hep-ph].

\bibitem{phi}
F. Archilli, D. Babusci, D. Badoni, et al., Phys. Lett., {\bf B706} (2012) 251.

\bibitem{pd}
A.V. Derbin, A.S. Kayunov, and V.N. Muratova, arxiv:1007.3387 [nucl-ex].

\bibitem{pd1}
G. Bellini, J. Benziger, D. Bick et al. (Borexino Collaboration),
arxiv:1203.6258 [hep-ex].

\bibitem{Mood}
J. E. Moody, F. Wilczek, Phys. Rev. (1984) {\bf D30} 130.

\bibitem{Dob}
B. A. Dobrescu and I. Mocioiu, JHEP {\bf 11} 005 (2006);

\bibitem{Alex}
Yu.A. Alexandrov, "Fundamental Properties of the Neutron", Moscow, Energoizdat, 1982, Second Edition;
and Oxford Univ. Press, N-Y, 1992.

\bibitem{Schw}
J. Schwinger, Phys. Rev. {\bf 73} (1948) 407.

\bibitem{Mug}
S.F. Mughabghab, "Atlas of Neutron Resonances: Resonance Parameters and Thermal Cross Sections. Z=1-100." (5th Ed.)

\bibitem{Bar}
A.L. Barabanov, V.R. Skoy, Nucl. Phys., {\bf A644} (1998) 54.

\bibitem{Lyu}
V.V. Lyuboshitz, V.L. Lyuboshitz, Yad. Fiz., {\bf  62} (1999) 1404.

\bibitem{Barb}
A.L. Barabanov, "Symmetries and spin-angular correlations in reactions and 
decays" (Fizmatlit, Moscow) 2010 (in Russian).

\bibitem{Vor}
V.V. Voronin, V.V. Fedorov, I.A. Kuznetsov, Pis'ma v ZhETF {\bf 90} (2009) 7.

\end{thebibliography}
\end{document}